\documentclass[%
 reprint,
 amsmath,amssymb,
 aps,
]{revtex4-1}

\usepackage{graphicx}% Include figure files
\usepackage{dcolumn}% Align table columns on decimal point
\usepackage{bm}% bold math
\usepackage{hyperref}% add hypertext capabilities
\usepackage{subfigure}
\usepackage[mathlines]{lineno}% Enable numbering of text and display math
\begin{document}

%\preprint{APS/123-QED}

\title{Electroweak and Top Results from ATLAS}%

\author{Marilyn Marx}
%\altaffiliation[Also at ]{Physics Department, XYZ University.}
\email{marx@cern.ch}
\affiliation{School of Physics and Astronomy, University of Manchester, Oxford Road, M13 9PL Manchester, UK}
\collaboration{On behalf of the ATLAS Collaboration}

%\affiliation{Second institution and/or address}
%\affiliation{Third institution}


\begin{abstract}
An overview of recent measurements of electroweak and top quark physics is given. In particular, total and differential diboson cross sections, limits on anomalous triple gauge couplings as well as top quark production cross sections and properties, such as charge asymmetry, top quark polarization and $Wtb$ vertex measurements, are presented. Proton-proton collision data produced at the LHC at $\sqrt{s}=7$~TeV and $\sqrt{s}=8$~TeV collected with the ATLAS detector are used.
\end{abstract}

\maketitle

%\tableofcontents
%Further information \url{http://authors.aps.org/revtex4/}.

\section{\label{sec:intro} Introduction}

Diboson and top quark production measurements at the Large Hadron Collider (LHC) allow to test the electroweak sector of the Standard Model (SM) to high precision. Diboson production is a significant and irreducible background to Higgs production and it is sensitive to the production and decay of new particles predicted in models with extended Higgs sectors, extra vector bosons, extra dimensions or models such as Supersymmetry and Technicolor. Furthermore, diboson production allows to probe, in a model-independent way and at unprobed energies, triple gauge boson couplings which are a fundamental prediction of the non-Abelian SU(2)$\times$U(1) gauge structure of electroweak theory. The top quark, produced in abundance at the LHC, is the heaviest fundamental SM particle and has a large coupling to the Higgs boson. Due to its extremely short lifetime, the top quark decays before it hadronizes and allows one to probe many interesting properties related to its production and decay.

The measurements presented here use proton-proton collision data produced at the LHC at centre of mass energies of $\sqrt{s}=7$~TeV and $\sqrt{s}=8$~TeV in 2011 and 2012, respectively, and collected with the ATLAS detector~\cite{ATLASJINST}. The integrated luminosities used for the various analyses are maximally 5 fb$^{-1}$ and 20 fb$^{-1}$ for 7~TeV and 8~TeV data, respectively, or subsets thereof.

\section{\label{sec:ew} Electroweak results}

\subsection{Diboson cross sections}

\begin{figure}[htbp]
\subfigure[\label{fig:Wg}]{\includegraphics[width=.4\textwidth]{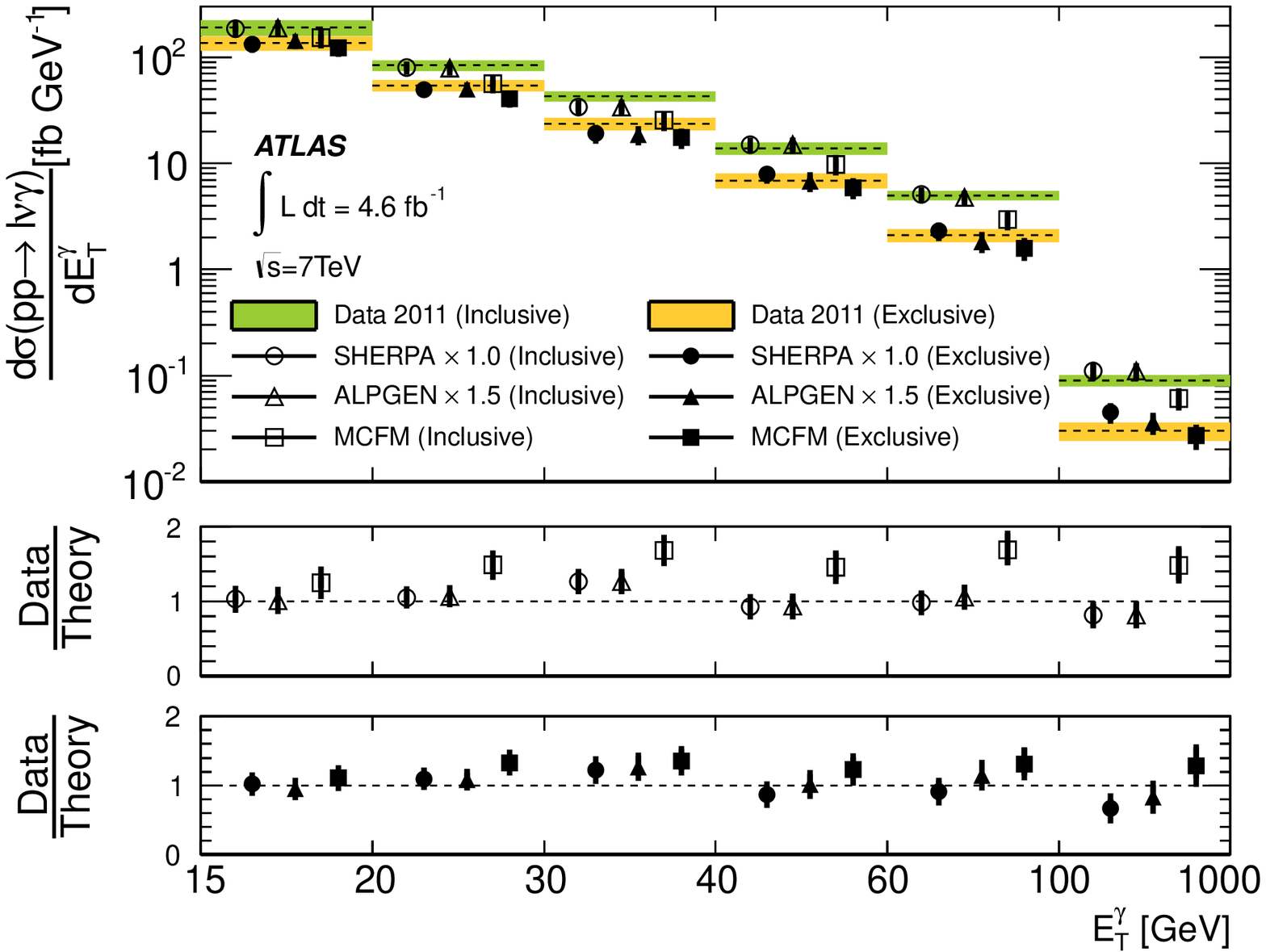}}
\subfigure[\label{fig:Zg}]{\includegraphics[width=.4\textwidth]{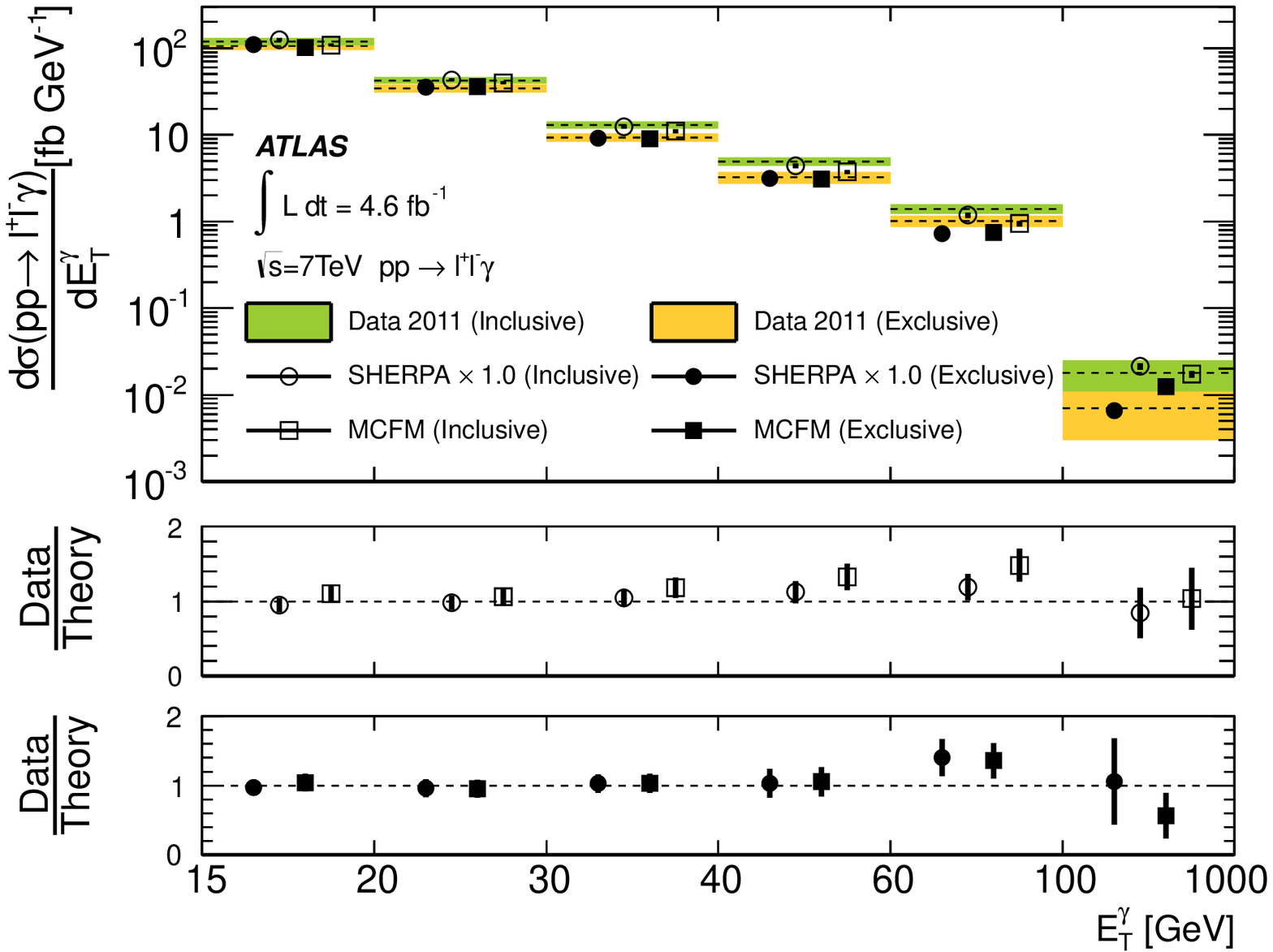}}
\caption{Inclusive and exclusive differential cross section measurements as a function of photon transverse energy for (a) $W(\to\ell^\pm\nu)\gamma$ and (b) $Z(\to\ell^+\ell^-)\gamma$ production~\cite{Aad:2013izg}.}
\label{fig:WZg}
\end{figure}

$W\gamma$ and $Z\gamma$ production cross sections are measured at 7~TeV as a function of the photon transverse energy $E^\gamma_{\rm T}$~\cite{Aad:2013izg}. Final states with leptonic decays of the massive gauge bosons are used. Both inclusive and exclusive cross section measurements are performed, where the latter requires a jet veto for all jets with $E_{\rm T}>30$~GeV. The data are unfolded to correct the measured values for detector effects such as limited acceptance, finite resolution and imperfect efficiency. The differential cross section results are shown in figure~\ref{fig:WZg}. Good agreement is seen with multi-leg leading order (LO) generators Alpgen~\cite{alpgen} and Sherpa~\cite{Sherpa} for the inclusive and exclusive descriptions. The next-to-leading order (NLO) MCFM~\cite{MCFM} inclusive prediction underestimates the data as multiple quark and gluon emission is not accounted for in the implementation. Normalised unfolded differential cross sections are also given as a function of jet multiplicity as well as diboson transverse invariant mass $m_{\rm T}^{W\gamma}$ or invariant mass $m^{Z\gamma}$ in the $W\gamma$ or $Z\gamma$ case, respectively. The latter two distributions are used to set limits on technicolor models.

The $W^+W^-$ total production cross section is measured using $\ell^+\nu\ell^-\nu$ final states using the full 7~TeV dataset~\cite{ATLAS:2012mec}. A jet veto is applied to suppress background contributions and represents the main systematic uncertainty in this measurement. The total $W^+W^-$ cross section is measured to be
\begin{equation}
\sigma_{WW}^{\rm tot}=51.9\pm2.0{\rm~(stat)}\pm3.9{\rm~(syst)}\pm2.0{\rm~(lumi)}\,{\rm pb}
\label{eq:WW7}
\end{equation}
which is compatible with a SM prediction of $\sigma_{WW}^{\rm NLO}=44.7^{+2.1}_{-1.9}\,{\rm pb}$. Normalised unfolded differential cross sections are measured as a function of the leading lepton transverse momentum $p_{\rm T}$ spectrum.

\begin{figure*}[htbp]
\subfigure[\label{fig:WZ7}]{\includegraphics[width=.4\textwidth]{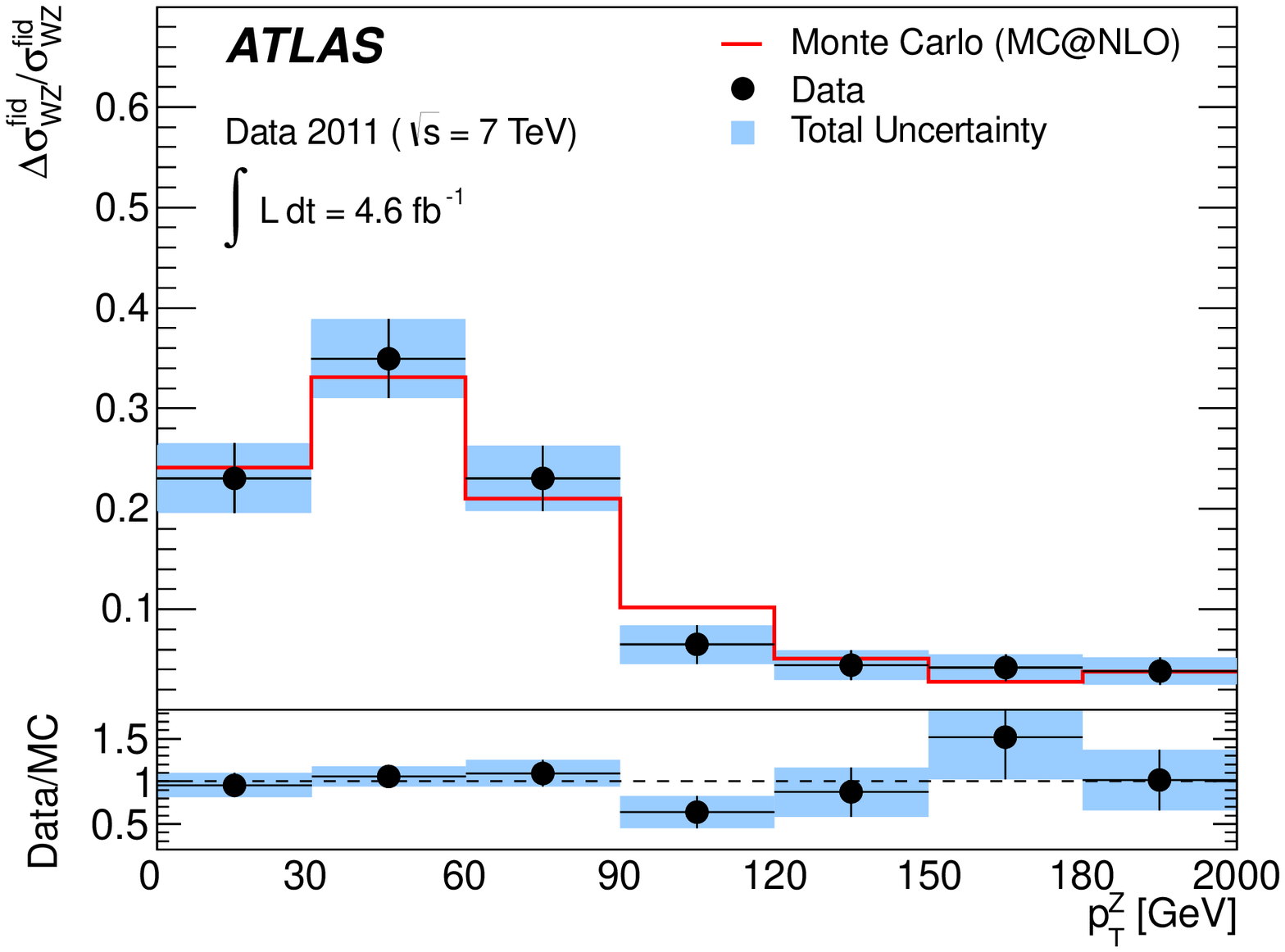}}
\subfigure[\label{fig:WZ8}]{\includegraphics[width=.4\textwidth]{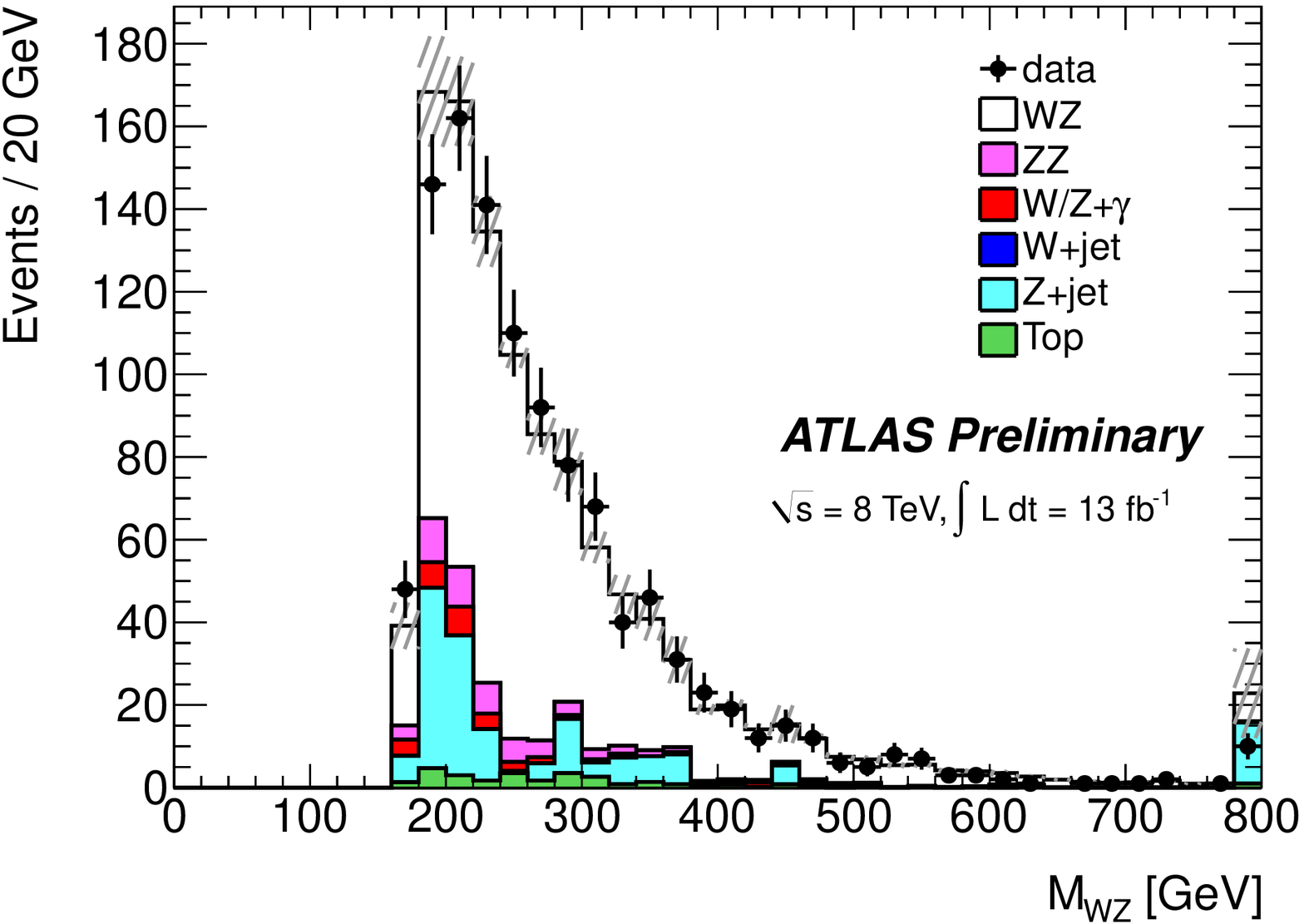}}
\caption{(a) The unfolded $p_{\rm T}^Z$ distribution in 7~TeV data~\cite{Aad:2012twa}. (b) The non-unfolded $m_{WZ}$ distribution in 8~TeV data~\cite{ATLAS-CONF-2013-021}.}
\end{figure*}

The $W^\pm Z$ production cross section is measured at 7~TeV~\cite{Aad:2012twa} and 8~TeV~\cite{ATLAS-CONF-2013-021} using final states with three charged leptons and large missing transverse momentum. The total production cross section results are
\begin{equation}
\sigma_{WZ}^{\rm tot}=19.0^{+1.4}_{-1.3}{\rm~(stat)}\pm0.9{\rm~(syst)}\pm0.4{\rm~(lumi)}\,{\rm pb}
\label{eq:WZ7}
\end{equation}
\begin{equation}
\sigma_{WZ}^{\rm tot}=20.3^{+0.8}_{-0.7}{\rm~(stat)}^{+1.2}_{-1.1}{\rm~(syst)}^{+0.7}_{-0.6}{\rm~(lumi)}\,{\rm pb}
\label{eq:WZ8}
\end{equation}
for 7~TeV and 8~TeV, respectively, and are in good agreement with SM predictions of $\sigma_{WZ}^{\rm NLO}=17.6^{+1.1}_{-1.0}\,{\rm pb}$ and $\sigma_{WZ}^{\rm NLO}=20.3\pm0.8\,{\rm pb}$, respectively. For 7~TeV data, normalised unfolded differential cross sections are measured as a function of the $Z$ boson $p_{\rm T}^Z$, shown in figure~\ref{fig:WZ7}, and as a function of the diboson invariant mass $m_{WZ}$. The non-unfolded $m_{WZ}$ distribution in 8~TeV data is shown in figure~\ref{fig:WZ8}.

\begin{figure}[htbp]
\includegraphics[width=.35\textwidth]{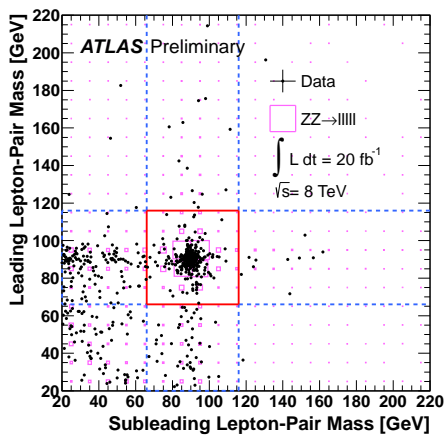}
\caption{The invariant mass distribution of the leading $Z$ boson candidate as a function of the invariant mass of the subleading $Z$ boson candidate~\cite{ATLAS-CONF-2013-020}.}
\label{fig:ZZ2D}
\end{figure}

The total $ZZ$ production cross section measurements are performed for 7~TeV~\cite{Aad:2012awa} and 8~TeV~\cite{ATLAS-CONF-2013-020} data. The results are 
\begin{equation}
\sigma_{ZZ}^{\rm tot}=6.7\pm0.7{\rm~(stat)}^{+0.4}_{-0.3}{\rm~(syst)}\pm0.3{\rm~(lumi)}\,{\rm pb}
\label{eq:ZZ7}
\end{equation}
\begin{equation}
\sigma_{ZZ}^{\rm tot}=7.1^{+0.5}_{-0.4}{\rm~(stat)}\pm0.3{\rm~(syst)}\pm0.2{\rm~(lumi)}\,{\rm pb}
\label{eq:ZZ8}
\end{equation}
for 7~TeV and 8~TeV in agreement with SM predictions of $\sigma_{ZZ}^{\rm NLO}=5.89^{+0.22}_{-0.18}\,{\rm pb}$ and $\sigma_{ZZ}^{\rm NLO}=7.2^{+0.3}_{-0.2}\,{\rm pb}$, respectively. For the 7~TeV measurement, $\ell^+\ell^-\ell^+\ell^-$ and $\ell^+\ell^-\nu\nu$ channels are used, the former can have off-shell bosons. Normalised unfolded differential cross sections are measured as a function of the leading $Z$ boson $p_{\rm T}$, the difference in azimuthal angle between the two leptons from the leading $Z$ boson decay $\Delta\phi(\ell\ell)$ and the diboson invariant mass $m^{ZZ}$ or transverse invariant mass $m^{ZZ}_{\rm T}$ for the $ZZ\to\ell^+\ell^-\ell^+\ell^-$ and $ZZ\to\ell^+\ell^-\nu\nu$ channels, respectively. For the 8~TeV result, only the $\ell^+\ell^-\ell^+\ell^-$ channel with on-shells bosons is considered. The mass distribution of the leading and subleading $Z$ boson candidates is shown in figure~\ref{fig:ZZ2D}.

\begin{figure}[htbp]
\includegraphics[width=.4\textwidth]{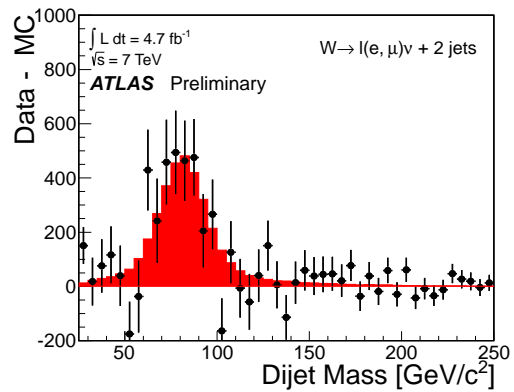}
\caption{The background subtracted dijet invariant mass distribution in $\ell\nu jj$ candidate events for the sum of electron and muon channels~\cite{ATLAS-CONF-2012-157}.}
\label{fig:WWWZmjj}
\end{figure}

As higher statistics datasets are recorded, it is possible to measure semileptonic diboson decays. The combined $WW/WZ\to\ell\nu jj$ production cross section is measured using the full 7~TeV dataset~\cite{ATLAS-CONF-2012-157}. This is a challenging measurement particularly at the LHC as the backgrounds grow much faster than the signal compared to the Tevatron. This means that the signal is dominated by large $W/Z+$ jets backgrounds. The cross section is extracted with a binned maximum likelihood fit to the dijet mass distribution $m_{jj}$ shown in figure~\ref{fig:WWWZmjj}. The observed signal significance is $3.3\sigma$. The measured cross section is
\begin{equation}
\sigma_{WW+WZ}^{\rm tot}=72\pm9{\rm~(stat)}\pm15{\rm~(syst)}\pm13{\rm~(MC~stat)}\,{\rm pb}
\label{eq:WWWZ7}
\end{equation}
which is in agreement with the SM prediction of $\sigma_{WW+WZ}^{\rm NLO}=63.4\pm2.6\,{\rm pb}$.

\subsection{Anomalous triple gauge coupling limits}

The model-independent effective Lagrangian for charged and neutral anomalous triple gauge boson couplings (aTGC) can be expressed as
\begin{widetext}
\begin{eqnarray}
\mathcal{L}_{WWV} &=&ig_{WWV}\left[{g^{V}_{1}}(W^{\dagger}_{\mu\nu}W^{\mu}V^{\nu}-W_{\mu\nu}W^{\dagger\mu}V^{\nu})+{\kappa^{V}}W^{\dagger}_{\mu}W_{\nu}V^{\mu\nu} +\frac{{\lambda^{V}}}{m^{2}_{W}}W^{\dagger}_{\rho\mu}W^{\mu}_{\nu}V^{\nu\rho}\right]\\
\mathcal{L}_{ZZV} &=&-\frac{e}{M^2_Z}\left[{f^{V}_{4}}(\partial_\mu V^{\mu\beta})Z_\alpha(\partial^\alpha Z_{\beta})+{f^{V}_{5}}(\partial^\sigma V_{\sigma\mu})\tilde{Z}^{\mu\beta}Z_{\beta})\right]\\
\mathcal{L}_{Z\gamma V} &=&-ie\left[{h^{V}_{3}}\tilde{F}^{\mu\nu}Z_\mu\frac{(\Box+m^2_V)}{m^2_Z}V_\nu+{h^{V}_{4}}\tilde{F}^{\mu\nu}Z^\alpha\frac{(\Box+m^2_V)}{m^4_Z}\partial_\alpha\partial_\mu V_\nu\right]
\end{eqnarray}
\end{widetext}
where $V=Z,\gamma$ and $V_{\mu\nu}=\partial_\mu V_\nu-\partial_\nu V_\mu$. The $WWV$ vertices are allowed in the SM whereas the $ZZV$ and $Z\gamma V$ vertices are not. In the SM, $g^V_1$ and $\kappa^V$ are equal to one whereas all other couplings vanish. The presence of aTGCs gives a change in the production rate and is visible in the distributions of different kinematic variables. Sensitivity is gained by using shape distributions to set one and two dimensional limits on the aTGC parameters. For $W\gamma/Z\gamma$ production the exclusive $E^\gamma_{\rm T}$ distribution is used to set limits whereas for $WW$ the leading lepton $p_{\rm T}$, for $WZ$ the $Z$ boson $p_{\rm T}$ and for $ZZ$ the leading $Z$ boson $p_{\rm T}$ distributions are used. The 95\% C.L. observed limits from the various channels are summarised in tables~\ref{tab:WWV}~--~\ref{tab:ZgV}. No deviations from the SM are observed.
\begin{table*}[htbp]
\begin{ruledtabular}
\begin{tabular}{lccccc}
Process & $\Delta g^Z_1$ & $\Delta\kappa_Z$& $\Delta\kappa_\gamma$ & $\lambda_Z$ &  $\lambda_\gamma$\\ 
\hline
$W^\pm\gamma$ & / & / & $[-0.41,0.46]$ & / & $[-0.065,0.061]$ \\
$W^+W^-$ & $[-0.039,0.0052]$ & $[-0.039,0.0052]$ & $=\cot^2\theta_W(\Delta g^Z_1-\Delta\kappa_Z)$ & $[-0.039,0.0052]$ & $=\lambda_Z$\\
$W^\pm Z$ & $[-0.057,0.093]$ & $[-0.37,0.57]$ & / & $[-0.046,0.047]$ & / \\
\end{tabular}
\end{ruledtabular}
\caption{Observed aTGC limits at the 95\% C.L. without a form factor from $W\gamma$, $WW$ and $WZ$ production~\cite{Aad:2013izg,ATLAS:2012mec,Aad:2012twa}.}
\label{tab:WWV}
\end{table*}
\begin{table}[htbp]
\begin{ruledtabular}
\begin{tabular}{lcccc}
Process & $f^\gamma_4\times10^2$ & $f^Z_4\times10^2$ & $f^\gamma_5\times10^2$ & $f^Z_5\times10^2$ \\ 
\hline
$ZZ$ & $[-1.5,1.5]$ & $[-1.3,1.3]$ & $[-1.6,1.5]$ & $[-1.3,1.3]$  \\
\end{tabular}
\end{ruledtabular}
\caption{Observed aTGC limits at the 95\% C.L. without a form factor from $ZZ$ production~\cite{Aad:2012awa}.}
\label{tab:ZZV}
\end{table}
\begin{table}[htbp]
\begin{ruledtabular}
\begin{tabular}{lcccc}
Process & $h^\gamma_3\times10^2$ & $h^Z_3\times10^2$ & $h^\gamma_4\times10^5$ & $h^Z_4\times10^5$ \\ 
\hline
$Z\gamma$ & $[-1.5,1.6]$ & $[-1.3,1.4]$ & $[-9.4,9.2]$ & $[-8.7,8.7]$  \\
\end{tabular}
\end{ruledtabular}
\caption{Observed aTGC limits at the 95\% C.L. without a form factor from $Z\gamma$ production~\cite{Aad:2013izg}.}
\label{tab:ZgV}
\end{table}

\section{\label{sec:top} Top results}

\subsection{Cross sections}

\begin{figure*}[htbp]
\subfigure[\label{fig:txsec}]{\includegraphics[width=.37\textwidth]{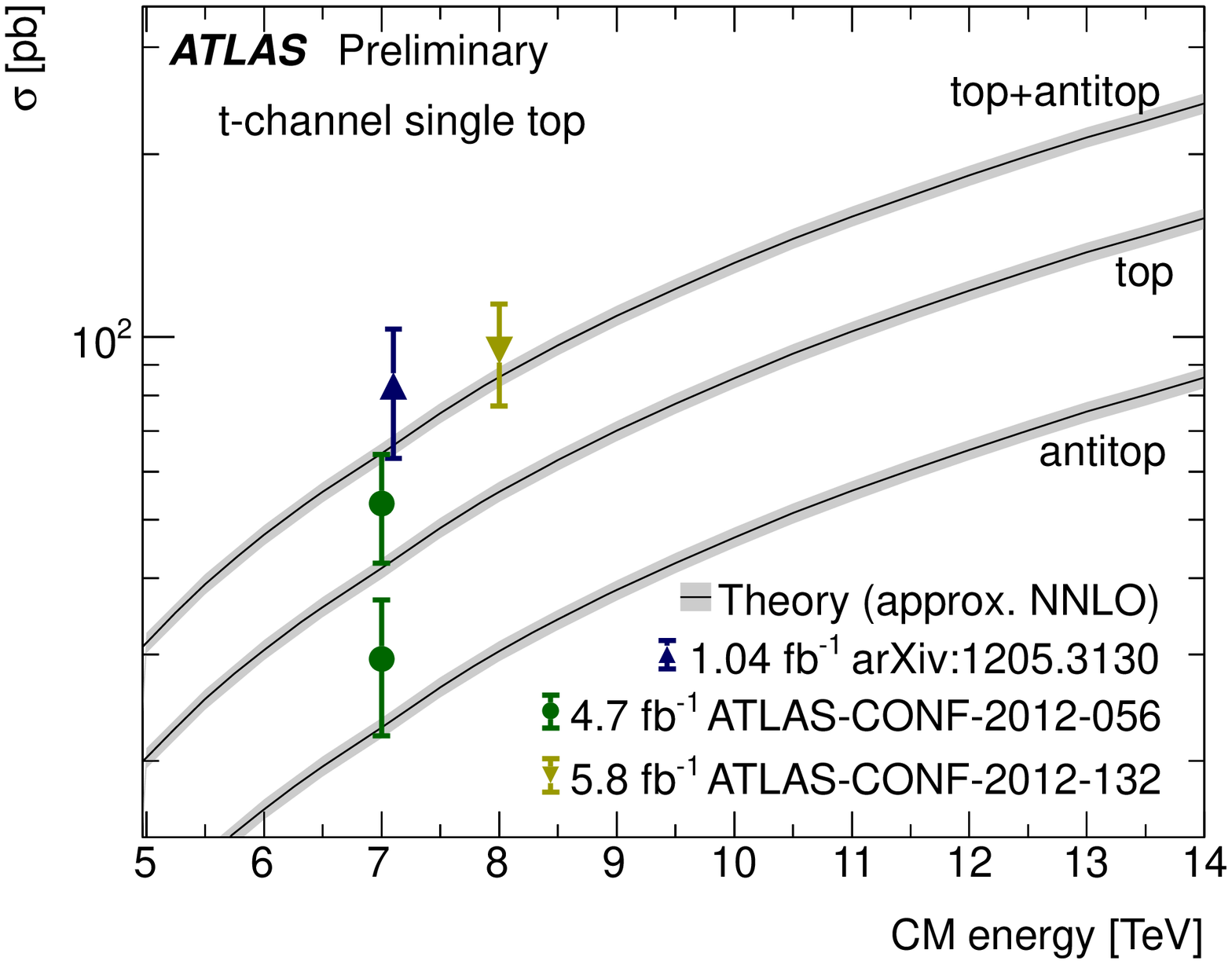}}
\subfigure[\label{fig:ttxsec}]{\includegraphics[width=.45\textwidth]{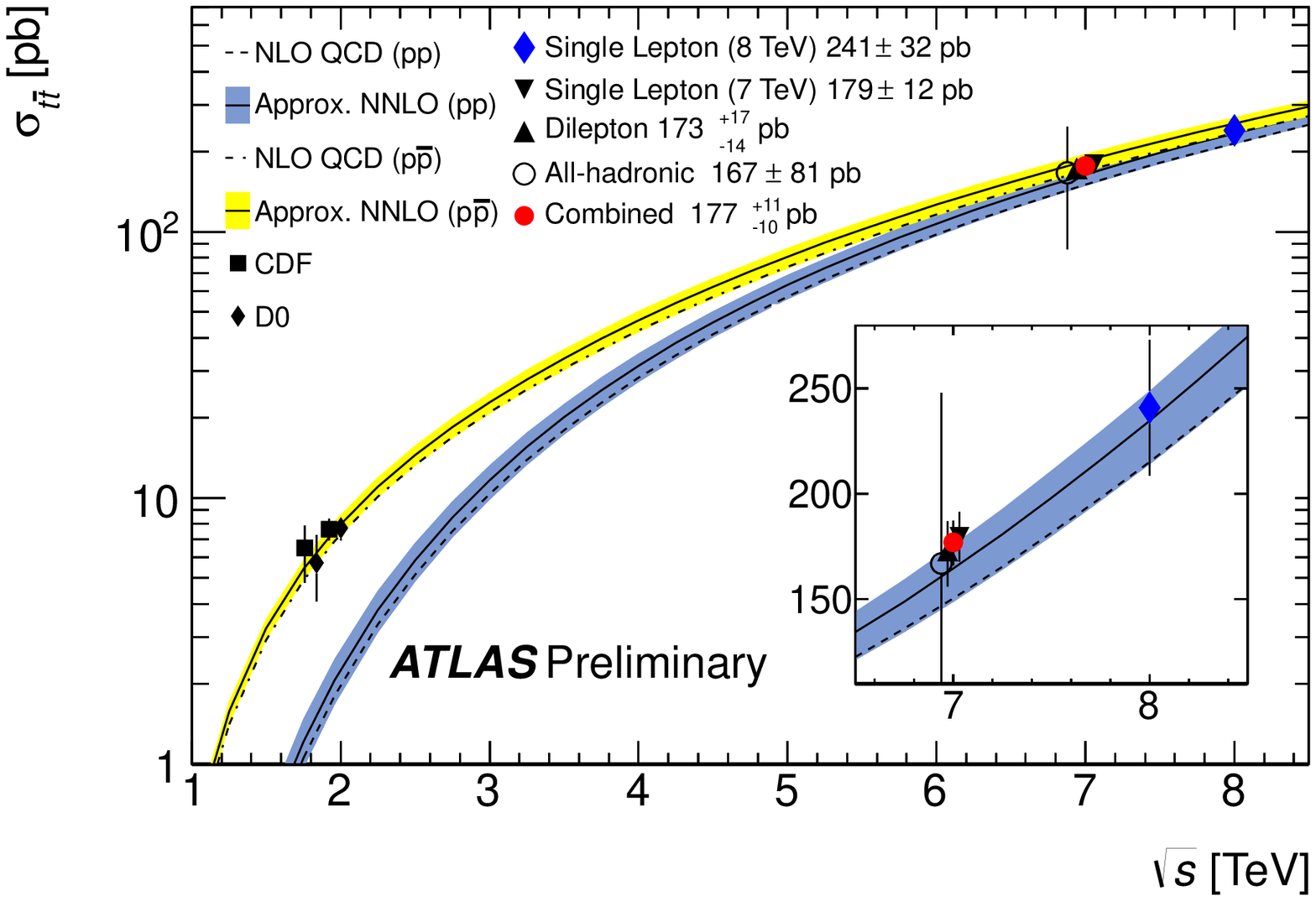}}
\caption{Measurements of (a) single top~\cite{ATLAS-CONF-2012-132} and (b) $t\bar{t}$~\cite{ATLAS-CONF-2012-149,summary} production cross sections as a function of centre of mass energy.}
\end{figure*}

At the LHC, single top quarks are produced through three main electroweak interactions, $t$-channel and $s$-channel production as well as production in association with a $W$ boson. An 8~TeV measurement of the single top production cross section is performed using the $t$-channel, which is the dominant process~\cite{ATLAS-CONF-2012-132}. Events with semileptonic top quark decays are considered. The cross section is extracted using a neural network based discriminant in the two and three jet bins. The main background contributions are from $W$+jets, QCD multijet and other top quark production. The measured cross section is
\begin{equation}
\sigma_{t}=95\pm2{\rm~(stat)}\pm18{\rm~(syst)}\,{\rm pb}
\end{equation}
which is in agreement with the approximate next-to-next-to-leading order (NNLO) prediction of $\sigma_{t}^{\rm NNLO}=87.8^{+3.4}_{-1.9}\,{\rm pb}$~\cite{Kidonakis:2011wy,Kidonakis:2012rm}. The measured and predicted cross sections can be seen as a function of $\sqrt{s}$ in figure~\ref{fig:txsec}. The coupling strength of the $Wtb$ vertex, accessible in this production channel, is measured using the ratio of the measured to the predicted cross sections. Assuming $|V_{tb}|\gg|V_{ts}|,|V_{td}|$, the limit on the CKM matrix element is $V_{tb}=1.04^{+0.10}_{-0.11}$ whereas assuming $V_{tb}\le1$ gives a 95\% C.L. lower limit of $|V_{tb}|>0.80$.

Top quark pairs ($t\bar{t}$)  are mainly ($\sim80\%$) produced through gluon-gluon fusion at the LHC. The $t\bar{t}$ inclusive production cross section is measured using 8~TeV data~\cite{ATLAS-CONF-2012-149}. Events where one $W$ boson decays leptonically and one hadronically are used. The analysis requires three or more jets with at least one $b$-tagged jet. A multivariate likelihood template fit is used to extract the cross section, which is shown as a function of $\sqrt{s}$ in figure~\ref{fig:ttxsec}. The 8~TeV measurement has a larger uncertainty compared to the 7~TeV measurements due to less aggressive Monte Carlo signal modelling uncertainties. The measured cross section is
\begin{equation}
\sigma_{t\bar{t}}=241\pm2{\rm~(stat)}\pm31{\rm~(syst)}\pm9{\rm~(lumi)}\,{\rm pb}
\end{equation}
compared to a predicted cross section at the approximate NNLO~\cite{Kidonakis:2010dk,Kidonakis:2012rm,HATHOR} of $\sigma_{t\bar{t}}^{\rm NNLO}=238^{+22}_{-24}\,{\rm pb}$.

\begin{figure}[htbp]
\includegraphics[width=.4\textwidth]{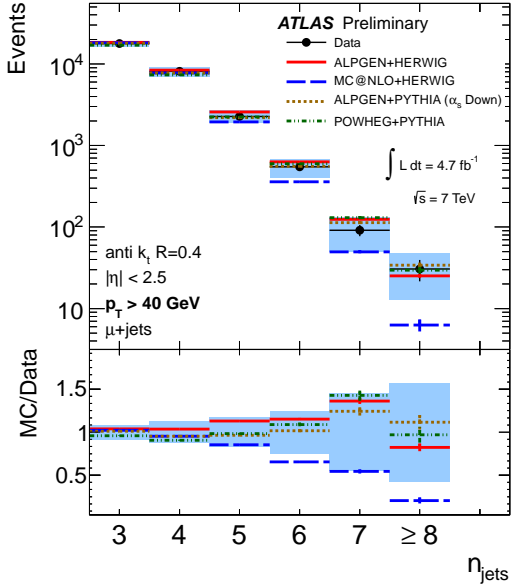}
\caption{Jet multiplicities for jet $p_{\rm T}$ values greater than 40~GeV in the muon channel. The data points, in black with statistical uncertainties, are compared to different MC predictions. The blue shaded bands represent the combined statistical and systematic uncertainties~\cite{ATLAS-CONF-2012-155}.}
\label{fig:njet}
\end{figure}

Unfolded differential $t\bar{t}$ cross sections are measured relative to the inclusive $t\bar{t}$ production cross section as a function of the invariant mass, the $p_{\rm T}$ and the rapidity of the $t\bar{t}$ system, which is fully reconstructed using a kinematic likelihood fit method~\cite{Aad:2012hg}. $t\bar{t}$ production is an important background for new searches. Semileptonic events with one lepton in the final state are selected. These measurements, which are dominated by systematic uncertainties, are useful because they are sensitive to wide resonances and QCD radiation. Using the same production channel, the unfolded jet multiplicity is measured for different jet $p_{\rm T}$ thresholds ranging from 25~GeV to 80~GeV~\cite{ATLAS-CONF-2012-155}. The distributions allow to constrain models of initial and final state radiation and to test perturbative QCD at the LHC. The jet multiplicities are shown in figure~\ref{fig:njet} for data and different simulations. Some generator simulations are disfavoured by the data as for example the MC@NLO prediction, which consistently estimates a lower number of jets.

\subsection{Properties}

\subsubsection{Charge asymmetry}

Top quark pair production is expected to be symmetric under charge conjugation at leading order in QCD. However, NLO corrections to the $q\bar{q}\to t\bar{t}$ process introduce small asymmetries in the rapidity $y$ distributions of the top quarks. Since the Tevatron was a $p$-$\bar{p}$ collider, this charge asymmetry was clearly visible as a forward-backward asymmetry as the $t$ quark is preferentially emitted along the $p$ direction and the $\bar{t}$ quark preferentially along the $\bar{p}$ direction. Both CDF~\cite{Aaltonen:2011kc} and D0~\cite{Abazov:2011rq} reported observed shifts that are larger than the expected shifts. At the LHC the problem is slightly different since the initial state is symmetric. Experimentally, this translates into a more forward $y$ distribution for the $t$ quark than for the $\bar{t}$ quark. This can be explained by the same fact as mentioned earlier that $t$ quarks are on average emitted mostly along the $q$ direction and that quarks, in the case of a proton, generally carry more momentum than antiquarks. Unless specific cuts are applied, the LHC is therefore more sensitive to the width of the $y$ distributions than to the mean. The charge asymmetry in $t\bar{t}$ events is defined as
\begin{equation}
A_C=\frac{1}{2}+\frac{N(\Delta|y|>0)-N(\Delta|y|<0)}{N(\Delta|y|>0)+N(\Delta|y|<0)}
\label{eq:Ac}
\end{equation}
where $\Delta|y|=|y_t|-|y_{\bar{t}}|$ is the difference in absolute top and antitop quark rapidities. To compute this $A_C$ fraction, the number of events with positive and negative $\Delta|y|$ values are measured and the charge of the quarks is deduced from the charge of the lepton produced in the $t$ or $\bar{t}$ quark decay. This has been measured in ATLAS in the single and dilepton channels. A combination of the two channels has been performed~\cite{ATLAS-CONF-2012-057} and the charge asymmetry factor is measured to be 
\begin{equation}
A_C=0.029\pm0.018{\rm~(stat)}\pm0.014{\rm~(syst)}
\end{equation}
which is compatible with the MC@NLO prediction of $A_C=0.006\pm0.002$~\cite{mcatnlo}.

\subsubsection{Top quark polarization}

\begin{figure}[htbp]
\includegraphics[width=.4\textwidth]{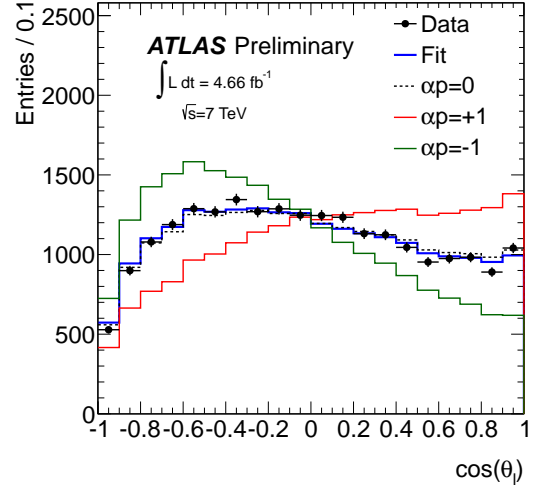}
\caption{The reconstructed $\cos\theta_l$ distribution showing the data and the best fit result as well as simulated distributions for zero, completely positive and negative polarizations in the electron plus jets channel~\cite{ATLAS-CONF-2012-133}.}
\label{fig:toppol}
\end{figure}

The short lifetime of the top quark allows one to measure its polarization directly from its decay products, which carry the full spin information that has not yet been degraded due to hadronization effects. The top quark polarization is measured for $t\bar{t}$ events in the lepton plus jets channel~\cite{ATLAS-CONF-2012-133}. The top quark pair is fully reconstructed using a likelihood method. The fraction of positively polarised top quarks can be written as
\begin{equation}
f_p=\frac{1}{2}+\frac{N(\cos\theta_l>0)-N(\cos\theta_l<0)}{N(\cos\theta_l>0)+N(\cos\theta_l<0)}.
\label{eq:fp}
\end{equation}
where $\theta_l$ is the polar lepton angle in the parent top quark frame. A template fit to the reconstructed $\cos\theta_l$ distribution in data, shown in figure~\ref{fig:toppol}, is used to extract this polarization fraction. The measured value is 
\begin{equation}
f_p=0.470\pm0.009{\rm~(stat)}^{+0.023}_{-0.032}{\rm~(syst)}
\end{equation}
which is in agreement with a SM expectation of unpolarised top quarks equivalent to $f_p=0.5$.

\subsubsection{$W$ boson polarization}

Top quark pair events can probe the $Wtb$ vertex which is defined by the $V-A$ structure of electroweak interactions. This decay vertex determines the helicity states of the produced $W$ boson, which can be in three different helicity states. The $F_0$, $F_L$ and $F_R$ helicity fractions measure the fraction of longitudinally polarised, left-handed or right-handed helicity states, respectively. The NNLO QCD predictions~\cite{helfracth} for these helicity fractions are 
\begin{eqnarray}
F_0&=&0.687\pm0.005\\
F_L&=&0.311\pm0.005\\
F_R&=&0.0017\pm0.0001. 
\end{eqnarray} 
Experimentally, these helicity fractions can be measured through template fits of the $\cos\theta^*$ distribution, where $\theta^*$ is the angle between the lepton from the $W$ boson decay and the reversed $b$ quark direction in the rest frame of the $W$ boson. ATLAS and CMS data produced at 7~TeV have been used to derive combined measurements of the $W$ boson helicity fractions~\cite{ATLAS-CONF-2013-033,Aad:2012ky}. The results are 
\begin{eqnarray}
 F_0&=&0.626\pm0.034{\rm~(stat)}\pm0.049{\rm~(syst)}\\
 F_L&=&0.359\pm0.021{\rm~(stat)}\pm0.028{\rm~(syst)}\\
 F_R&=&0.015\pm0.034{\rm~(stat+syst)}
\end{eqnarray} 
where $F_R$ is derived from the $F_0$ and $F_L$ measurements by assuming that $F_0+F_L+F_R=1$.

\begin{figure}[htbp]
\includegraphics[width=.48\textwidth]{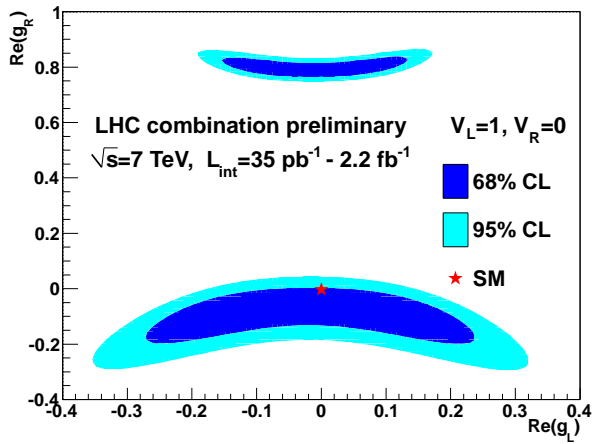}
\caption{Allowed 68\% and 95\% C.L. regions of the anomalous $Wtb$ coupling parameters assuming $V_L=1$ and $V_R=0$~\cite{AguilarSaavedra:2006fy,AguilarSaavedra:2008zc,ATLAS-CONF-2013-033}.}
\label{fig:Wpol}
\end{figure}

This measurement can also be used to set limits on anomalous $Wtb$ couplings using an effective Lagrangian approach. The form of the Lagrangian for the $Wtb$ vertex is~\cite{AguilarSaavedra:2006fy}
\begin{widetext}
\begin{equation}
\mathcal{L}_{Wtb} =\frac{g}{\sqrt{2}}\bar{b}\left[\gamma^\mu(V_LP_L+V_RP_R)+\frac{i\sigma^{\mu\nu q_\nu}}{m_W}(g_LP_L+g_RP_R)\right]tW^-_\mu+h.c.
\end{equation}
\end{widetext}
where $V_L=V_{tb}\approx1$ and $V_R=0$ while $g_L$ and $g_R$, the anomalous coupling parameters on which the limits will be set, vanish at tree level in the SM. Assuming that the imaginary parts of all couplings are zero, the 68\% and 95\% C.L. limits are set on the real part of $g_L$ and $g_R$, ${\rm Re}(g_L)$ and ${\rm Re}(g_R)$, and the two dimensional representation of these limits can be seen in figure~\ref{fig:Wpol}. The region around $Re(g_R)$ is not excluded by the data used in this analysis but it is disfavoured by single top cross section measurements.

\subsubsection{Search for CP violation}

\begin{figure}[htbp]
\includegraphics[width=.48\textwidth]{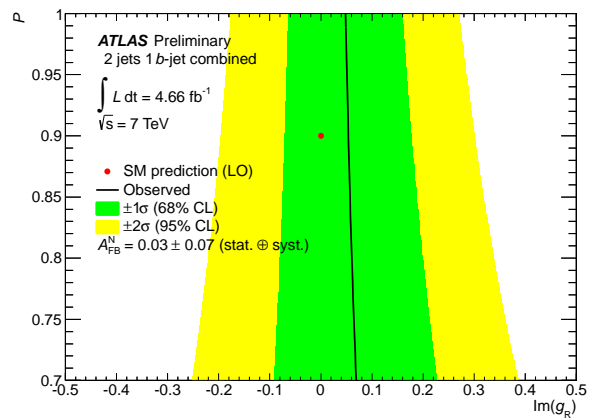}
\caption{Constraints on ${\rm Im}(g_R)$ as a function of the top quark polarization~\cite{AguilarSaavedra:2010nx,ATLAS-CONF-2013-032}.}
\label{fig:CP}
\end{figure}

A further test of the $Wtb$ vertex is possible through a search for $CP$ violation using the single top $t$-channel, where the top quarks are expected to be highly polarised, with lepton plus jets final states~\cite{AguilarSaavedra:2010nx,ATLAS-CONF-2013-032}. A forward-backward asymmetry factor, $A_{\rm FB}$, with respect to the normal to the plane defined by the $W$ boson momentum and the top quark polarization direction is measured to be
\begin{equation}
A_{\rm FB}=0.31\pm0.065{\rm~(stat)}^{+0.029}_{-0.031}{\rm~(syst)}
\label{eq:AFB}
\end{equation}
which is consistent with $CP$ invariance for which $A_{\rm FB}=0$ is expected. The result is used to set the first experimental limits on the imaginary part of the anomalous coupling parameter $g_R$. The 68\% and 95\% C.L. limits on ${\rm Im}(g_R)$ are shown in figure~\ref{fig:CP} as a function of the top quark polarization. If a top quark polarization of 0.9, which is the theoretical prediction, is assumed, the observed limits on ${\rm Im}(g_R)$ at the 95\% C.L. are $[-0.20,0.30]$ in agreement with the SM.

\section{\label{sec:summ} Summary}

Recent electroweak and top measurements, based on 7~TeV and 8~TeV data collected with the ATLAS detector in 2011 and 2012 respectively, have been presented. Fiducial and total diboson production cross sections have been measured and are mostly dominated by systematic uncertainties. Limits on aTGC parameters have been set in many channels and surpass the precision of the Tevatron results. The first normalised unfolded differential diboson cross sections have been measured and are statistically dominated. Single top and $t\bar{t}$ production cross sections have been measured using 8~TeV data and differential $t\bar{t}$ cross sections have been measured using 7~TeV data, all of which are systematically dominated. Precision measurements of different properties related to the production and decay of the top quark have been presented. No significant deviations from SM predictions have been found in any of these measurements, many of which still need to be repeated with the full 8~TeV datatset. The presented measurements are crucial milestones for the understanding of Higgs boson production and searches for new physics.

%\begin{acknowledgments}
%We wish to acknowledge the support of the author community in using REV\TeX{}, offering suggestions and encouragement, testing new versions,\dots.
%\end{acknowledgments}

%\nocite{*}
%\pagebreak
\bibliography{apssamp}% Produces the bibliography via BibTeX.

\providecommand{\noopsort}[1]{}\providecommand{\singleletter}[1]{#1}%
\begin{thebibliography}{32}%
\makeatletter
\providecommand \@ifxundefined [1]{%
 \@ifx{#1\undefined}
}%
\providecommand \@ifnum [1]{%
 \ifnum #1\expandafter \@firstoftwo
 \else \expandafter \@secondoftwo
 \fi
}%
\providecommand \@ifx [1]{%
 \ifx #1\expandafter \@firstoftwo
 \else \expandafter \@secondoftwo
 \fi
}%
\providecommand \natexlab [1]{#1}%
\providecommand \enquote  [1]{``#1''}%
\providecommand \bibnamefont  [1]{#1}%
\providecommand \bibfnamefont [1]{#1}%
\providecommand \citenamefont [1]{#1}%
\providecommand \href@noop [0]{\@secondoftwo}%
\providecommand \href [0]{\begingroup \@sanitize@url \@href}%
\providecommand \@href[1]{\@@startlink{#1}\@@href}%
\providecommand \@@href[1]{\endgroup#1\@@endlink}%
\providecommand \@sanitize@url [0]{\catcode `\\12\catcode `\$12\catcode
  `\&12\catcode `\#12\catcode `\^12\catcode `\_12\catcode `\%12\relax}%
\providecommand \@@startlink[1]{}%
\providecommand \@@endlink[0]{}%
\providecommand \url  [0]{\begingroup\@sanitize@url \@url }%
\providecommand \@url [1]{\endgroup\@href {#1}{\urlprefix }}%
\providecommand \urlprefix  [0]{URL }%
\providecommand \Eprint [0]{\href }%
\providecommand \doibase [0]{http://dx.doi.org/}%
\providecommand \selectlanguage [0]{\@gobble}%
\providecommand \bibinfo  [0]{\@secondoftwo}%
\providecommand \bibfield  [0]{\@secondoftwo}%
\providecommand \translation [1]{[#1]}%
\providecommand \BibitemOpen [0]{}%
\providecommand \bibitemStop [0]{}%
\providecommand \bibitemNoStop [0]{.\EOS\space}%
\providecommand \EOS [0]{\spacefactor3000\relax}%
\providecommand \BibitemShut  [1]{\csname bibitem#1\endcsname}%
\let\auto@bib@innerbib\@empty
%</preamble>
\bibitem [{\citenamefont {{ATLAS Collaboration}}(2008)}]{ATLASJINST}%
  \BibitemOpen
  \bibfield  {author} {\bibinfo {author} {\bibnamefont {{ATLAS
  Collaboration}}},\ }\href {\doibase 10.1088/1748-0221/3/08/S08003} {\bibfield
   {journal} {\bibinfo  {journal} {JINST}\ }\textbf {\bibinfo {volume} {3}},\
  \bibinfo {pages} {S08003} (\bibinfo {year} {2008})}\BibitemShut {NoStop}%
\bibitem [{\citenamefont {{ATLAS
  Collaboration}}(2013{\natexlab{a}})}]{Aad:2013izg}%
  \BibitemOpen
  \bibfield  {author} {\bibinfo {author} {\bibnamefont {{ATLAS
  Collaboration}}},\ }\href@noop {} {\  (\bibinfo {year}
  {2013}{\natexlab{a}})},\ \Eprint {http://arxiv.org/abs/1302.1283}
  {arXiv:1302.1283 [hep-ex]} \BibitemShut {NoStop}%
%%CITATION = ARXIV:1302.1283;%%
\bibitem [{\citenamefont {Mangano}\ \emph {et~al.}(2003)\citenamefont
  {Mangano}, \citenamefont {Piccinini}, \citenamefont {Polosa}, \citenamefont
  {Moretti},\ and\ \citenamefont {Pittau}}]{alpgen}%
  \BibitemOpen
  \bibfield  {author} {\bibinfo {author} {\bibfnamefont {M.~L.}\ \bibnamefont
  {Mangano}}, \bibinfo {author} {\bibfnamefont {F.}~\bibnamefont {Piccinini}},
  \bibinfo {author} {\bibfnamefont {A.}~\bibnamefont {Polosa}}, \bibinfo
  {author} {\bibfnamefont {M.}~\bibnamefont {Moretti}}, \ and\ \bibinfo
  {author} {\bibfnamefont {R.}~\bibnamefont {Pittau}},\ }\href@noop {}
  {\bibfield  {journal} {\bibinfo  {journal} {JHEP}\ }\textbf {\bibinfo
  {volume} {07}},\ \bibinfo {pages} {001} (\bibinfo {year} {2003})},\ \Eprint
  {http://arxiv.org/abs/hep-ph/0206293} {arXiv:hep-ph/0206293} \BibitemShut
  {NoStop}%
\bibitem [{\citenamefont {Gleisberg}\ \emph {et~al.}(2009)\citenamefont
  {Gleisberg}, \citenamefont {H{\"o}che}, \citenamefont {Krauss}, \citenamefont
  {Sch{\"o}nherr}, \citenamefont {Schumann} \emph {et~al.}}]{Sherpa}%
  \BibitemOpen
  \bibfield  {author} {\bibinfo {author} {\bibfnamefont {T.}~\bibnamefont
  {Gleisberg}}, \bibinfo {author} {\bibfnamefont {S.}~\bibnamefont
  {H{\"o}che}}, \bibinfo {author} {\bibfnamefont {F.}~\bibnamefont {Krauss}},
  \bibinfo {author} {\bibfnamefont {M.}~\bibnamefont {Sch{\"o}nherr}}, \bibinfo
  {author} {\bibfnamefont {S.}~\bibnamefont {Schumann}},  \emph {et~al.},\
  }\href {\doibase 10.1088/1126-6708/2009/02/007} {\bibfield  {journal}
  {\bibinfo  {journal} {JHEP}\ }\textbf {\bibinfo {volume} {0902}},\ \bibinfo
  {pages} {007} (\bibinfo {year} {2009})},\ \Eprint
  {http://arxiv.org/abs/0811.4622} {arXiv:0811.4622 [hep-ph]} \BibitemShut
  {NoStop}%
%%CITATION = ARXIV:0811.4622;%%
\bibitem [{\citenamefont {Campbell}\ \emph {et~al.}(2011)\citenamefont
  {Campbell}, \citenamefont {Ellis},\ and\ \citenamefont {Williams}}]{MCFM}%
  \BibitemOpen
  \bibfield  {author} {\bibinfo {author} {\bibfnamefont {J.~M.}\ \bibnamefont
  {Campbell}}, \bibinfo {author} {\bibfnamefont {R.~K.}\ \bibnamefont {Ellis}},
  \ and\ \bibinfo {author} {\bibfnamefont {C.}~\bibnamefont {Williams}},\
  }\href@noop {} {\bibfield  {journal} {\bibinfo  {journal} {JHEP}\ }\textbf
  {\bibinfo {volume} {1107}},\ \bibinfo {pages} {018} (\bibinfo {year}
  {2011})},\ \Eprint {http://arxiv.org/abs/1105.0020} {arXiv:1105.0020
  [hep-ph]} \BibitemShut {NoStop}%
%%CITATION = ARXIV:1105.0020;%%
\bibitem [{\citenamefont {{ATLAS
  Collaboration}}(2012{\natexlab{a}})}]{ATLAS:2012mec}%
  \BibitemOpen
  \bibfield  {author} {\bibinfo {author} {\bibnamefont {{ATLAS
  Collaboration}}},\ }\href@noop {} {\  (\bibinfo {year}
  {2012}{\natexlab{a}})},\ \Eprint {http://arxiv.org/abs/1210.2979}
  {arXiv:1210.2979 [hep-ex]} \BibitemShut {NoStop}%
%%CITATION = ARXIV:1210.2979;%%
\bibitem [{\citenamefont {{ATLAS
  Collaboration}}(2012{\natexlab{b}})}]{Aad:2012twa}%
  \BibitemOpen
  \bibfield  {author} {\bibinfo {author} {\bibnamefont {{ATLAS
  Collaboration}}},\ }\href {\doibase 10.1140/epjc/s10052-012-2173-0}
  {\bibfield  {journal} {\bibinfo  {journal} {Eur.Phys.J.}\ }\textbf {\bibinfo
  {volume} {C72}},\ \bibinfo {pages} {2173} (\bibinfo {year}
  {2012}{\natexlab{b}})},\ \Eprint {http://arxiv.org/abs/1208.1390}
  {arXiv:1208.1390 [hep-ex]} \BibitemShut {NoStop}%
%%CITATION = ARXIV:1208.1390;%%
\bibitem [{\citenamefont {{ATLAS
  Collaboration}}(2013{\natexlab{b}})}]{ATLAS-CONF-2013-021}%
  \BibitemOpen
  \bibfield  {author} {\bibinfo {author} {\bibnamefont {{ATLAS
  Collaboration}}},\ }\ \bibinfo {number} {ATLAS-CONF-2013-021}\ (\bibinfo
  {year} {2013})\ \bibinfo {note}
  {\url{https://cds.cern.ch/record/1525557}}\BibitemShut {NoStop}%
\bibitem [{\citenamefont {{ATLAS
  Collaboration}}(2013{\natexlab{c}})}]{Aad:2012awa}%
  \BibitemOpen
  \bibfield  {author} {\bibinfo {author} {\bibnamefont {{ATLAS
  Collaboration}}},\ }\href {\doibase 10.1007/JHEP03(2013)128} {\bibfield
  {journal} {\bibinfo  {journal} {JHEP}\ }\textbf {\bibinfo {volume} {1303}},\
  \bibinfo {pages} {128} (\bibinfo {year} {2013}{\natexlab{c}})},\ \Eprint
  {http://arxiv.org/abs/1211.6096} {arXiv:1211.6096 [hep-ex]} \BibitemShut
  {NoStop}%
%%CITATION = ARXIV:1211.6096;%%
\bibitem [{\citenamefont {{ATLAS
  Collaboration}}(2013{\natexlab{d}})}]{ATLAS-CONF-2013-020}%
  \BibitemOpen
  \bibfield  {author} {\bibinfo {author} {\bibnamefont {{ATLAS
  Collaboration}}},\ }\ \bibinfo {number} {ATLAS-CONF-2013-020}\ (\bibinfo
  {year} {2013})\ \bibinfo {note}
  {\url{https://cds.cern.ch/record/1525555}}\BibitemShut {NoStop}%
\bibitem [{\citenamefont {{ATLAS
  Collaboration}}(2012{\natexlab{c}})}]{ATLAS-CONF-2012-157}%
  \BibitemOpen
  \bibfield  {author} {\bibinfo {author} {\bibnamefont {{ATLAS
  Collaboration}}},\ }\ \bibinfo {number} {ATLAS-CONF-2012-157}\ (\bibinfo
  {year} {2012})\ \bibinfo {note}
  {\url{https://cds.cern.ch/record/1493586}}\BibitemShut {NoStop}%
\bibitem [{\citenamefont {{ATLAS
  Collaboration}}(2012{\natexlab{d}})}]{ATLAS-CONF-2012-132}%
  \BibitemOpen
  \bibfield  {author} {\bibinfo {author} {\bibnamefont {{ATLAS
  Collaboration}}},\ }\ \bibinfo {number} {ATLAS-CONF-2012-132}\ (\bibinfo
  {year} {2012})\ \bibinfo {note}
  {\url{https://cds.cern.ch/record/1478371}}\BibitemShut {NoStop}%
\bibitem [{\citenamefont {Kidonakis}(2011)}]{Kidonakis:2011wy}%
  \BibitemOpen
  \bibfield  {author} {\bibinfo {author} {\bibfnamefont {N.}~\bibnamefont
  {Kidonakis}},\ }\href {\doibase 10.1103/PhysRevD.83.091503} {\bibfield
  {journal} {\bibinfo  {journal} {Phys.Rev.}\ }\textbf {\bibinfo {volume}
  {D83}},\ \bibinfo {pages} {091503} (\bibinfo {year} {2011})},\ \Eprint
  {http://arxiv.org/abs/1103.2792} {arXiv:1103.2792 [hep-ph]} \BibitemShut
  {NoStop}%
%%CITATION = ARXIV:1103.2792;%%
\bibitem [{\citenamefont {Kidonakis}(2012)}]{Kidonakis:2012rm}%
  \BibitemOpen
  \bibfield  {author} {\bibinfo {author} {\bibfnamefont {N.}~\bibnamefont
  {Kidonakis}},\ }\href@noop {} {\  (\bibinfo {year} {2012})},\ \Eprint
  {http://arxiv.org/abs/1210.7813} {arXiv:1210.7813 [hep-ph]} \BibitemShut
  {NoStop}%
%%CITATION = ARXIV:1210.7813;%%
\bibitem [{\citenamefont {{ATLAS
  Collaboration}}(2012{\natexlab{e}})}]{ATLAS-CONF-2012-149}%
  \BibitemOpen
  \bibfield  {author} {\bibinfo {author} {\bibnamefont {{ATLAS
  Collaboration}}},\ }\ \bibinfo {number} {ATLAS-CONF-2012-149}\ (\bibinfo
  {year} {2012})\ \bibinfo {note}
  {\url{https://cds.cern.ch/record/1493488}}\BibitemShut {NoStop}%
\bibitem [{\citenamefont {Kidonakis}(2010)}]{Kidonakis:2010dk}%
  \BibitemOpen
  \bibfield  {author} {\bibinfo {author} {\bibfnamefont {N.}~\bibnamefont
  {Kidonakis}},\ }\href {\doibase 10.1103/PhysRevD.82.114030} {\bibfield
  {journal} {\bibinfo  {journal} {Phys.Rev.}\ }\textbf {\bibinfo {volume}
  {D82}},\ \bibinfo {pages} {114030} (\bibinfo {year} {2010})},\ \Eprint
  {http://arxiv.org/abs/1009.4935} {arXiv:1009.4935 [hep-ph]} \BibitemShut
  {NoStop}%
%%CITATION = ARXIV:1009.4935;%%
\bibitem [{\citenamefont {Aliev}\ \emph {et~al.}(2011)\citenamefont {Aliev},
  \citenamefont {Lacker}, \citenamefont {Langenfeld}, \citenamefont {Moch},
  \citenamefont {Uwer} \emph {et~al.}}]{HATHOR}%
  \BibitemOpen
  \bibfield  {author} {\bibinfo {author} {\bibfnamefont {M.}~\bibnamefont
  {Aliev}}, \bibinfo {author} {\bibfnamefont {H.}~\bibnamefont {Lacker}},
  \bibinfo {author} {\bibfnamefont {U.}~\bibnamefont {Langenfeld}}, \bibinfo
  {author} {\bibfnamefont {S.}~\bibnamefont {Moch}}, \bibinfo {author}
  {\bibfnamefont {P.}~\bibnamefont {Uwer}},  \emph {et~al.},\ }\href {\doibase
  10.1016/j.cpc.2010.12.040} {\bibfield  {journal} {\bibinfo  {journal}
  {Comput.Phys.Commun.}\ }\textbf {\bibinfo {volume} {182}},\ \bibinfo {pages}
  {1034} (\bibinfo {year} {2011})},\ \Eprint {http://arxiv.org/abs/1007.1327}
  {arXiv:1007.1327 [hep-ph]} \BibitemShut {NoStop}%
%%CITATION = ARXIV:1007.1327;%%
\bibitem [{sum()}]{summary}%
  \BibitemOpen
  \href@noop {} {}\bibinfo {howpublished}
  {\href{https://twiki.cern.ch/twiki/bin/view/AtlasPublic/CombinedSummaryPlots}{ATLAS
  Public Results -- Physics Summary Plots}}\BibitemShut {NoStop}%
\bibitem [{\citenamefont {{ATLAS
  Collaboration}}(2013{\natexlab{e}})}]{Aad:2012hg}%
  \BibitemOpen
  \bibfield  {author} {\bibinfo {author} {\bibnamefont {{ATLAS
  Collaboration}}},\ }\href {\doibase 10.1140/epjc/s10052-012-2261-1}
  {\bibfield  {journal} {\bibinfo  {journal} {Eur.Phys.J.}\ }\textbf {\bibinfo
  {volume} {C73}},\ \bibinfo {pages} {2261} (\bibinfo {year}
  {2013}{\natexlab{e}})},\ \Eprint {http://arxiv.org/abs/1207.5644}
  {arXiv:1207.5644 [hep-ex]} \BibitemShut {NoStop}%
%%CITATION = ARXIV:1207.5644;%%
\bibitem [{\citenamefont {{ATLAS
  Collaboration}}(2012{\natexlab{f}})}]{ATLAS-CONF-2012-155}%
  \BibitemOpen
  \bibfield  {author} {\bibinfo {author} {\bibnamefont {{ATLAS
  Collaboration}}},\ }\ \bibinfo {number} {ATLAS-CONF-2012-155}\ (\bibinfo
  {year} {2012})\ \bibinfo {note}
  {\url{https://cds.cern.ch/record/1493494}}\BibitemShut {NoStop}%
\bibitem [{\citenamefont {{CDF Collaboration}}(2011)}]{Aaltonen:2011kc}%
  \BibitemOpen
  \bibfield  {author} {\bibinfo {author} {\bibnamefont {{CDF Collaboration}}},\
  }\href {\doibase 10.1103/PhysRevD.83.112003} {\bibfield  {journal} {\bibinfo
  {journal} {Phys.Rev.}\ }\textbf {\bibinfo {volume} {D83}},\ \bibinfo {pages}
  {112003} (\bibinfo {year} {2011})},\ \Eprint {http://arxiv.org/abs/1101.0034}
  {arXiv:1101.0034 [hep-ex]} \BibitemShut {NoStop}%
%%CITATION = ARXIV:1101.0034;%%
\bibitem [{\citenamefont {{D0 Collaboration}}(2011)}]{Abazov:2011rq}%
  \BibitemOpen
  \bibfield  {author} {\bibinfo {author} {\bibnamefont {{D0 Collaboration}}},\
  }\href {\doibase 10.1103/PhysRevD.84.112005} {\bibfield  {journal} {\bibinfo
  {journal} {Phys.Rev.}\ }\textbf {\bibinfo {volume} {D84}},\ \bibinfo {pages}
  {112005} (\bibinfo {year} {2011})},\ \Eprint {http://arxiv.org/abs/1107.4995}
  {arXiv:1107.4995 [hep-ex]} \BibitemShut {NoStop}%
%%CITATION = ARXIV:1107.4995;%%
\bibitem [{\citenamefont {{ATLAS
  Collaboration}}(2012{\natexlab{g}})}]{ATLAS-CONF-2012-057}%
  \BibitemOpen
  \bibfield  {author} {\bibinfo {author} {\bibnamefont {{ATLAS
  Collaboration}}},\ }\ \bibinfo {number} {ATLAS-CONF-2012-057}\ (\bibinfo
  {year} {2012})\ \bibinfo {note}
  {\url{https://cds.cern.ch/record/1453785}}\BibitemShut {NoStop}%
\bibitem [{\citenamefont {Frixione}\ and\ \citenamefont
  {Webber}(2002)}]{mcatnlo}%
  \BibitemOpen
  \bibfield  {author} {\bibinfo {author} {\bibfnamefont {S.}~\bibnamefont
  {Frixione}}\ and\ \bibinfo {author} {\bibfnamefont {B.~R.}\ \bibnamefont
  {Webber}},\ }\href@noop {} {\bibfield  {journal} {\bibinfo  {journal} {JHEP}\
  }\textbf {\bibinfo {volume} {0206}},\ \bibinfo {pages} {029} (\bibinfo {year}
  {2002})},\ \Eprint {http://arxiv.org/abs/hep-ph/0204244}
  {arXiv:hep-ph/0204244 [hep-ph]} \BibitemShut {NoStop}%
%%CITATION = HEP-PH/0204244;%%
\bibitem [{\citenamefont {{ATLAS
  Collaboration}}(2012{\natexlab{h}})}]{ATLAS-CONF-2012-133}%
  \BibitemOpen
  \bibfield  {author} {\bibinfo {author} {\bibnamefont {{ATLAS
  Collaboration}}},\ }\ \bibinfo {number} {ATLAS-CONF-2012-133}\ (\bibinfo
  {year} {2012})\ \bibinfo {note}
  {\url{https://cds.cern.ch/record/1478373}}\BibitemShut {NoStop}%
\bibitem [{\citenamefont {Czarnecki}\ \emph {et~al.}(2010)\citenamefont
  {Czarnecki}, \citenamefont {K\"orner},\ and\ \citenamefont
  {Piclum}}]{helfracth}%
  \BibitemOpen
  \bibfield  {author} {\bibinfo {author} {\bibfnamefont {A.}~\bibnamefont
  {Czarnecki}}, \bibinfo {author} {\bibfnamefont {J.~G.}\ \bibnamefont
  {K\"orner}}, \ and\ \bibinfo {author} {\bibfnamefont {J.~H.}\ \bibnamefont
  {Piclum}},\ }\href {\doibase 10.1103/PhysRevD.81.111503} {\bibfield
  {journal} {\bibinfo  {journal} {Phys. Rev. D}\ }\textbf {\bibinfo {volume}
  {81}},\ \bibinfo {pages} {111503} (\bibinfo {year} {2010})}\BibitemShut
  {NoStop}%
\bibitem [{\citenamefont {{ATLAS and CMS
  Collaborations}}(2013)}]{ATLAS-CONF-2013-033}%
  \BibitemOpen
  \bibfield  {author} {\bibinfo {author} {\bibnamefont {{ATLAS and CMS
  Collaborations}}},\ }\ \bibinfo {number} {ATLAS-CONF-2013-033}\ (\bibinfo
  {year} {2013})\ \bibinfo {note}
  {\url{https://cds.cern.ch/record/1527531}}\BibitemShut {NoStop}%
\bibitem [{\citenamefont {{ATLAS
  Collaboration}}(2012{\natexlab{i}})}]{Aad:2012ky}%
  \BibitemOpen
  \bibfield  {author} {\bibinfo {author} {\bibnamefont {{ATLAS
  Collaboration}}},\ }\href {\doibase 10.1007/JHEP06(2012)088} {\bibfield
  {journal} {\bibinfo  {journal} {JHEP}\ }\textbf {\bibinfo {volume} {1206}},\
  \bibinfo {pages} {088} (\bibinfo {year} {2012}{\natexlab{i}})},\ \Eprint
  {http://arxiv.org/abs/1205.2484} {arXiv:1205.2484 [hep-ex]} \BibitemShut
  {NoStop}%
%%CITATION = ARXIV:1205.2484;%%
\bibitem [{\citenamefont {Aguilar-Saavedra}\ \emph {et~al.}(2007)\citenamefont
  {Aguilar-Saavedra}, \citenamefont {Carvalho}, \citenamefont {Castro},
  \citenamefont {Veloso},\ and\ \citenamefont
  {Onofre}}]{AguilarSaavedra:2006fy}%
  \BibitemOpen
  \bibfield  {author} {\bibinfo {author} {\bibfnamefont {J.}~\bibnamefont
  {Aguilar-Saavedra}}, \bibinfo {author} {\bibfnamefont {J.}~\bibnamefont
  {Carvalho}}, \bibinfo {author} {\bibfnamefont {N.~F.}\ \bibnamefont
  {Castro}}, \bibinfo {author} {\bibfnamefont {F.}~\bibnamefont {Veloso}}, \
  and\ \bibinfo {author} {\bibfnamefont {A.}~\bibnamefont {Onofre}},\ }\href
  {\doibase 10.1140/epjc/s10052-007-0289-4} {\bibfield  {journal} {\bibinfo
  {journal} {Eur.Phys.J.}\ }\textbf {\bibinfo {volume} {C50}},\ \bibinfo
  {pages} {519} (\bibinfo {year} {2007})},\ \Eprint
  {http://arxiv.org/abs/hep-ph/0605190} {arXiv:hep-ph/0605190 [hep-ph]}
  \BibitemShut {NoStop}%
%%CITATION = HEP-PH/0605190;%%
\bibitem [{\citenamefont {Aguilar-Saavedra}(2009)}]{AguilarSaavedra:2008zc}%
  \BibitemOpen
  \bibfield  {author} {\bibinfo {author} {\bibfnamefont {J.}~\bibnamefont
  {Aguilar-Saavedra}},\ }\href {\doibase 10.1016/j.nuclphysb.2008.12.012}
  {\bibfield  {journal} {\bibinfo  {journal} {Nucl.Phys.}\ }\textbf {\bibinfo
  {volume} {B812}},\ \bibinfo {pages} {181} (\bibinfo {year} {2009})},\ \Eprint
  {http://arxiv.org/abs/0811.3842} {arXiv:0811.3842 [hep-ph]} \BibitemShut
  {NoStop}%
%%CITATION = ARXIV:0811.3842;%%
\bibitem [{\citenamefont {Aguilar-Saavedra}\ and\ \citenamefont
  {Bernabeu}(2010)}]{AguilarSaavedra:2010nx}%
  \BibitemOpen
  \bibfield  {author} {\bibinfo {author} {\bibfnamefont {J.}~\bibnamefont
  {Aguilar-Saavedra}}\ and\ \bibinfo {author} {\bibfnamefont {J.}~\bibnamefont
  {Bernabeu}},\ }\href {\doibase 10.1016/j.nuclphysb.2010.07.012} {\bibfield
  {journal} {\bibinfo  {journal} {Nucl.Phys.}\ }\textbf {\bibinfo {volume}
  {B840}},\ \bibinfo {pages} {349} (\bibinfo {year} {2010})},\ \Eprint
  {http://arxiv.org/abs/1005.5382} {arXiv:1005.5382 [hep-ph]} \BibitemShut
  {NoStop}%
%%CITATION = ARXIV:1005.5382;%%
\bibitem [{\citenamefont {{ATLAS
  Collaboration}}(2013{\natexlab{f}})}]{ATLAS-CONF-2013-032}%
  \BibitemOpen
  \bibfield  {author} {\bibinfo {author} {\bibnamefont {{ATLAS
  Collaboration}}},\ }\ \bibinfo {number} {ATLAS-CONF-2013-032}\ (\bibinfo
  {year} {2013})\ \bibinfo {note}
  {\url{https://cds.cern.ch/record/1527128}}\BibitemShut {NoStop}%
\end{thebibliography}%

%\input{bibliography}
%to add titles back, add
  %new.block.comma
  %link.open
  %format.btitle
  %"title" output.check
  %link.shut
 %after line 2947 in .bst

\end{document}